\documentclass{PoS}
\usepackage{amsmath}

\title{Nuclear medium modifications of properties of kaons measured around threshold with FOPI}

\ShortTitle{Kaons in medium with FOPI}

\author{\speaker{K. Piasecki (FOPI Collaboration)
}\thanks{}\\
        University of Warsaw, Faculty of Physics\\
        E-mail: \email{krzysztof.piasecki@fuw.edu.pl}}

\abstract{We report on the investigation of modifications 
of basic properties of K$^+$ and K$^-$ mesons emitted from 
collisions of Ni+Ni at beam energy of 1.91A~GeV. 
Experimental K$^- \slash \text{K}^+$ ratio are presented 
in a wide range of phase space parameterized by kinetic energy
and emission angle in the nucleon-nucleon centre of mass.
The $v_1$ component of the azimuthal distribution was extracted 
as a function of rapidity and transverse momentum for central, 
and semi-peripheral collisions.
A comparison of these patterns with the HSD transport model
favours the existence of the kaon-nucleon in-medium potential.
For the IQMD model, this interaction scenario is confirmed
in case of K$^-$, whereas for K$^+$ the picture is less clear.}

\FullConference{55th International Winter Meeting on Nuclear Physics\\
		23-27 January, 2017\\
		Bormio, Italy}

\begin{document}

\section{Introduction}

Effects of modifications of basic properties of hadrons 
(like mass and decay constant) propagating through the hot 
and/or dense nuclear medium are the subject of intensive studies 
throughout last 30 years~\cite{Kap,Bro,Scha,Lutz}. 
In particular, various theoretical approaches predict a somewhat 
repulsive interaction of K$^+$ and K$^0_s$ meson with nucleon, 
whereas the interaction of antikaon with nucleon is predicted 
to be strongly attractive. 
Historically, the kaon-nucleon interaction was first represented 
in terms of an additional potential $U_\text{KN}$ (positive/negative
for repulsion/attraction, respectively)~\cite{Scha}. 
However, more recent approaches based on the chiral effective field 
theory with coupled channels predict that the in-medium spectral function 
of antikaon gains significant width and becomes momentum-dependent~\cite{Lutz}. 
While these modifications cannot be measured directly, they are
predicted to alter observables like kinetic energy spectrum or flow 
(anisotropy of azimuthal distribution with respect to the reaction plane). 
During the collision of heavy ions the baryonic matter is found 
to exhibit strong flow effects. If a kaon (antikaon) propagates 
through such a medium, it is expected to be repelled from 
(resp. attracted to) the centers of density, and therefore, 
its flow pattern deviates (resp. gets similar to) that of 
the bulk of baryonic matter. 
On the other hand, as kaons leave the collision zone, 
their spectral functions have to return to the vacuum form. 
If the antikaon leaves the collision zone, and its effective mass 
is to increase, the simplest way to acquire extra energy is to decelerate.
For the kaon, the effect should be the opposite.

In spite of the importance of the strangeness production to 
the understanding of the nucleus-nucleus collisions 
around threshold energies, the experimental knowledge on the 
in-medium effects of kaons has been quite limited. 
The ratios of $\text{K}^- \slash \text{K}^+$ as function of kinetic energy,
compared to the predictions of the (R)BUU transport models~\cite{Wisn,Laue} 
or HSD models~\cite{Gasi} were strongly suggesting the non-zero $U_{\text{KN}}$ 
potentials, however, these ratios were obtained 
for narrow windows of the polar emission angle~\cite{Wisn,Laue},
and momentum~\cite{Gasi}. 
The comparisons of experimentally obtained flow patterns of K$^+$ emission
to the predictions of the (R)BUU transport model~\cite{Shin,Croc} were also 
displaying preference for the in-medium effects, 
however, again, the data were investigated experimentally 
in narrow windows of phase space, and the data volume allowed to obtain
results with significant uncertainties.
In addition, throughout the last decade the yield of the important
feeding of the $\phi$ meson decays into the emitted negative kaons was 
experimentally constrained to be $18 \pm 3$~\%~\cite{Agak,Pia6}. 
While the energy spectrum of negative kaons originating from 
decays of $\phi$ mesons seems to be considerably 
softer than that of K$^-$ emitted directly from the collision 
zone~\cite{Lore,Gasi}, the production and decays of $\phi$ mesons 
within the abovementioned transport models was either 
not implemented yet, or did not match the experimental findings. 

In addition, a very recent prediction of the UrQMD transport model 
calculations on the kinetic energy profile of K$^- \slash \text{K}^+$ 
ratio from Al+Al collisions at 1.9A GeV was found to nicely agree 
with the experimental profile~\cite{Blei}, despite the fact
that no in-medium modifications of kaon properties are implemented 
within this model. However, the K$^-$ and K$^+$ energy spectra compared 
separately exhibit some excess with respect to the experiment. 

Therefore more precise data, covering wider range of phase space 
appear to be needed to clarify the situation. 
Data presented in this proceedings paper aims to bring the problem 
a step closer to the solution.

\section{Experiment}

The S325 experiment was performed by the FOPI Collaboration 
at the SIS18 accelerator at GSI, Darmstadt. 
$^{58}$Ni ions were accelerated to the beam
kinetic energy of 1.91$A$ GeV and collided with the $^{nat}$Ni target. 

The target was surrounded by the central drift chamber (CDC) 
covering the polar angles of $27^{\circ} < \theta_\text{lab} < 113^{\circ}$.
The CDC was encircled by two Time-of-Flight detection systems, the
plastic scintillation barrel (PSB, $55^\circ < \theta_\text{lab} < 110^\circ$),
and the MMRPC barrel ($30^\circ < \theta_\text{lab} < 53^\circ$)~\cite{Kis}.
This set of detectors was mounted inside the magnet solenoid delivering
the magnetic field of \mbox{0.617 T}. 
A wide geometrical acceptance range of the setup,
matched with the good (PSB), and excellent (MMRPC) ToF resolution 
translated into the broad acceptance of phase space available 
within the back hemisphere with respect to the nucleon-nucleon
(NN) centre of mass.
The trigger based on the multiplicity of charged particles~\cite{Gobb,Siko} 
selected events ranging from most central to semi-peripheral 
collisions ($\sigma ~\approx~ 56\% ~\sigma_\text{geom}$). 
More details on the experimental setup and the identification 
of kaons can be found in~\cite{Ziny,Pia5}.

\section{Directed flow of charged kaons}

\begin{figure}[tb]
 \includegraphics[width=1\textwidth]{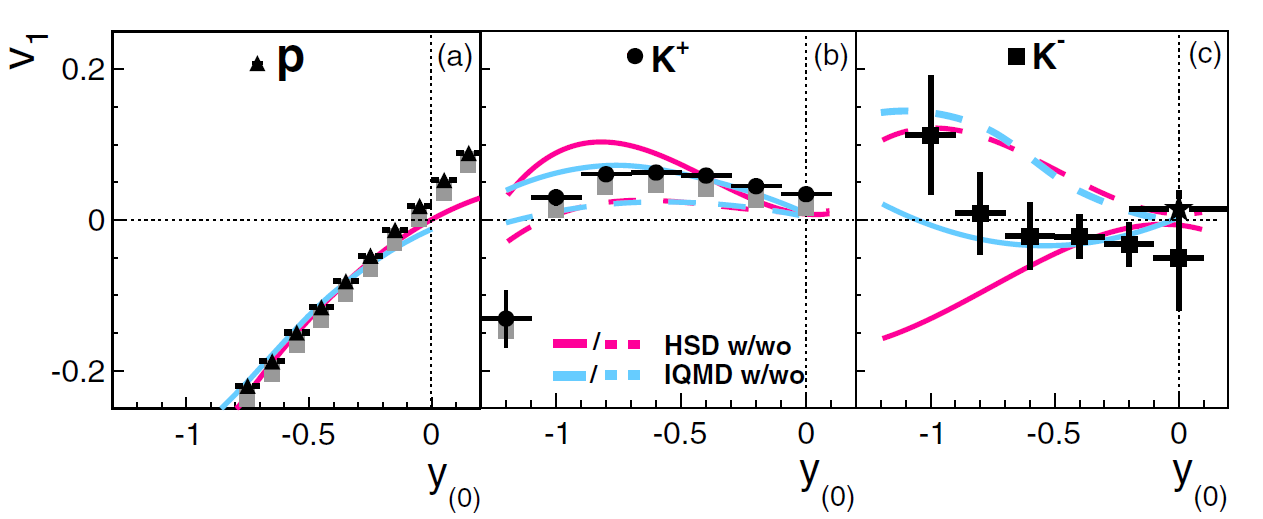}
 \caption{Directed flow ($v_1$) as function of reduced rapidity
 for protons, and charged kaons from collisions of Ni (1.91A GeV) + Ni.
 Lines depict predictions of the HSD (magenta), and IQMD (blue)
 transport models with (solid), and without (dashed) the in-medium effects. 
 See text for details. Picture taken from~\cite{Ziny}.}
 \label{fig1}
\end{figure}

In this section we summarize the analysis published in~\cite{Ziny}.
The azimuthal distribution of emitted particles with respect 
to the reaction plane is often decomposed in terms of the 
Fourier series, 

\begin{equation}
 \frac{dN}{d\phi} = \sum\limits_{i=1}^\infty v_\text{n} ~cos(n\phi) 
\end{equation}

\noindent where $v_\text{n}$ are the weights of the subsequent 
harmonic terms. 

The dependence of the $v_1$ parameter ("directed flow") of protons 
and charged kaons on the reduced rapidity in the NN reference frame
($y_{(0)} = (y_\text{Lab} - y_\text{NN}) / y_\text{NN}$), 
is shown in Fig.~\ref{fig1}. While for most rapidities the K$^+$ mesons
exhibit positive directed flow (for the backward hemisphere), 
in case of K$^-$ mesons the values of $v_1$ for the majority 
of data points are found to be either consistent with 0 
or slightly negative. 
These results were compared to the predictions of two transport models, 
IQMD~\cite{IQMD} and HSD~\cite{HSD}, as shown in Fig.~\ref{fig1} with 
blue, and magenta curves, respectively. The predictions where the in-medium
effects were switched off, are depicted by dashed lines. 
For the IQMD model predictions with the in-medium modifications, 
the following potentials were chosen:
$U_{\text{K}^{+}\text{N}} (\rho = \rho_0, p = 0) = +20$~MeV, and 
$U_{\text{K}^{-}\text{N}} (\rho = \rho_0, p = 0) = -45$~MeV,
where $\rho_0$ is the normal nuclear density. 
It seems, that the calculations assuming the in-medium effects
nicely reproduce the $v_1 (y_{(0)})$ pattern, 
although the choice of slightly smaller values of both $U_{\text{KN}}$ 
may give a better agreement with the experimental data. 
Within the HSD model, only the K$^+$ in-medium effects are parameterized
by the $U_{\text{KN}}$ potential. For the current comparison, 
it was set to +20~MeV. The K$^-$ mesons are treated as off-shell 
particles within the G-Matrix formalism, and therefore 
the potential picture should be perceived only as an approximation. 
For K$^-$ the strength of in-medium effects
corresponds to $U_{\text{K}^{-}\text{N}} (\rho = \rho_0, p = 0) = -50$~MeV. 
The model predictions obtained with the abovementioned potentials appear 
to generate too strong effect with respect to the experimental findings, 
and indicate that calculations with $U_\text{KN}$ about halved 
could be a better choice. 

\begin{figure}[tb]
 \begin{center}
 \includegraphics[width=0.65\textwidth]{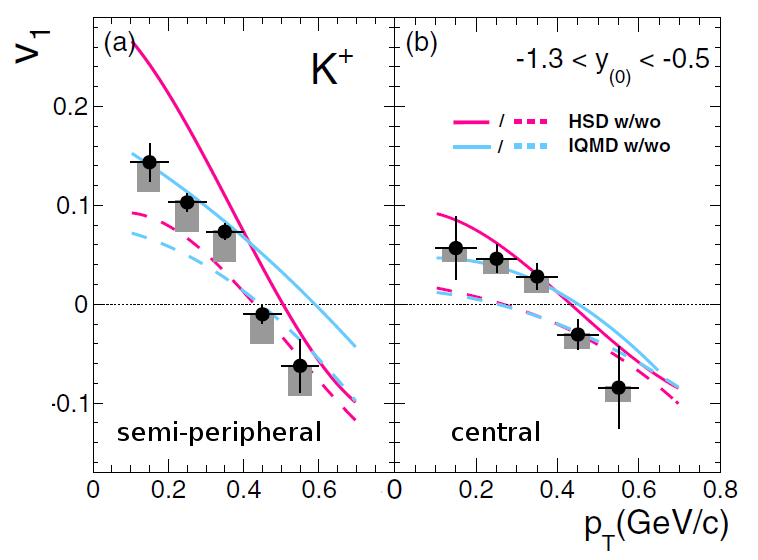}
 \caption{
 Directed flow ($v_1$) of K$^+$ mesons as function of transverse momentum for 
 (a) peripheral, and (b) central collisions of Ni (1.91A GeV) + Ni. 
 Lines depict predictions of the HSD (magenta), and IQMD (blue)
 transport models with (solid), and without (dashed) the in-medium effects. 
 See text for details. Picture taken from~\cite{Ziny}.}
 \label{fig2}
 \end{center}
\end{figure}

Fig.~\ref{fig2} shows the dependence of directed flow on transverse momentum
for the K$^+$ mesons measured within $-1.3 < y_{(0)} < -0.5$ range, 
for semi-peripheral, and central Ni+Ni collisions. 
The comparisons of calculations of the HSD, and IQMD transport models 
do not yield straightforward conclusions. 
For the semi-peripheral collisions the IQMD predictions do not seem 
to favour any of the two scenarios.
In case of the in-medium version of the HSD calculations, 
the predicted $v_1$ pattern at low $p_\text{T}$ exhibits much stronger
flow than that found experimentally, suggesting again, 
that the size of potential was too high. 
For the central collisions, the data points at low $p_\text{T}$
turned out to be situated between two scenarios, sugesting again, 
that about halved value of the tested potential could yield a better 
agreement. At higher $p_\text{T}$ the comparison is inconclusive.

\section{Phase space distribution of K$^- \slash \text{K}^+$ ratio}

\begin{figure}[tb]
 \begin{center}
  \includegraphics[width=0.9\textwidth]{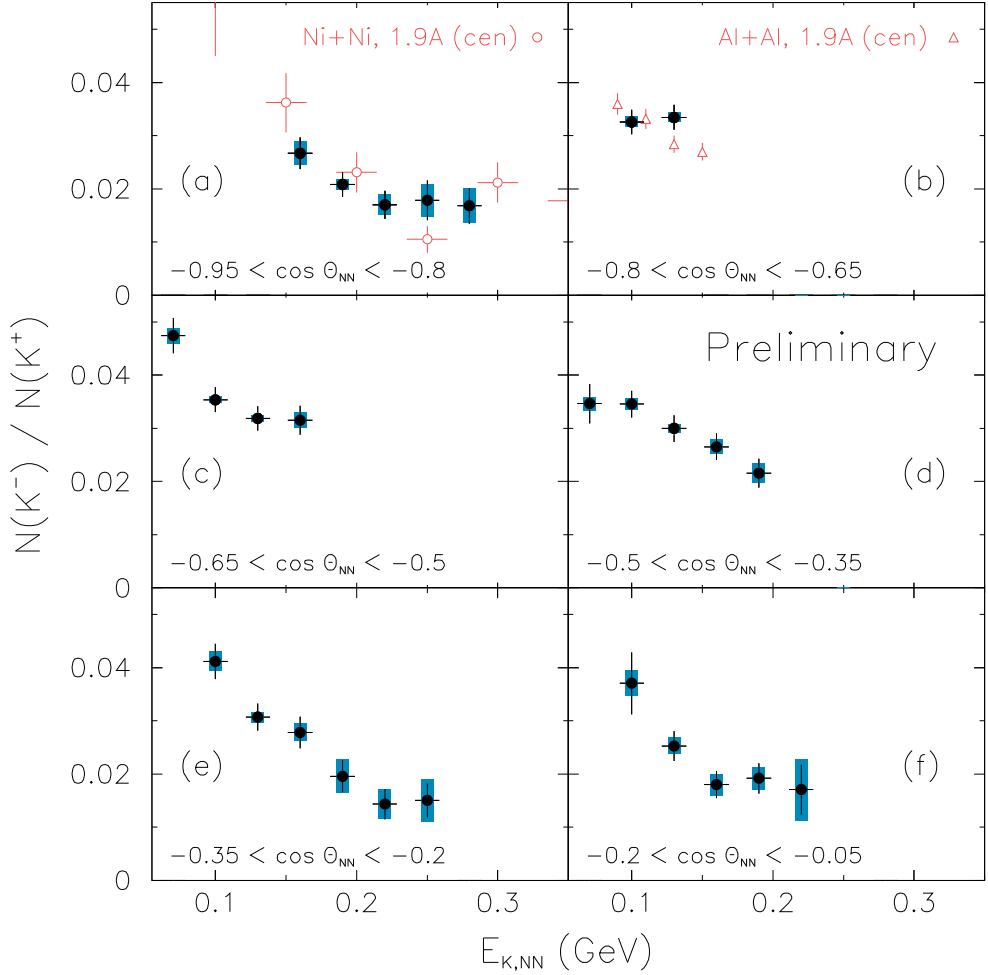}
  \caption{(Full circles:) ratios of K$^- \slash \text{K}^+$ as function of kinetic 
  energy for subsequent slices of $\cos ~\theta_\text{NN}$, from s325 experiment.
  (Open circles:) data for central Ni+Ni collisions, from previous
                  experiment, published in~\cite{Wisn}.
  (Open triangles:) data for central Al+Al collisions, published in~\cite{Gasi}.
  See text for details.}
  \label{fig3}
 \end{center}
\end{figure}

The reconstructed phase space yield of K$^-$ mesons was divided 
by that of positively charged kaons. In Fig.~\ref{fig3} we show for the 
first time this ratio as a function of kinetic energy in the NN reference frame
for subsequent slices of cos $\theta_\text{NN}$, 
spanning nearly all the emission angles in the backward hemisphere. 
The data points for the K$^- \slash \text{K}^+$ ratio 
from the central Ni+Ni collisions at the same beam energy, 
reported in~\cite{Wisn} are shown for rough comparison 
in panel (a) of this figure. 
The latter data were measured only within 
$-0.97 < \cos \theta_\text{NN} < -0.87$ range, and exhibit clearly larger 
statistical uncertainties than these obtained in the recent experiment.
Also, four data points are available for the collisions of Al+Al at
the same beam energy~\cite{Gasi}. As they were measured for the range
$-0.87 < \cos \theta_\text{NN} < -0.72$, they were placed 
in panel (b) for rough comparison. 
The systematic errors of our data points have been evaluated at the
1$\sigma$ confidence level, and include variations in the background
subtraction procedure, and track quality cuts. 
Although preliminary, these data are in the final stage of analysis,
and the theorists interested to compare this data to the prediction
of the transport model are invited to contact us.

\section{Summary}

The collisions of Ni+Ni at the beam kinetic energy of 1.91A GeV
were measured by the FOPI Collaboration.
The excellent time resolution of the MMRPC detector allowed to 
obtain more precise data on the charged kaon production 
in terms of the directed flow and the $\text{K}^- \slash \text{K}^+$ ratio 
in a wide acceptance. 
Comparisons of these distributions to the predictions of HSD and IQMD
transport models allows to get insight into the scale of 
modifications of kaon properties in the nucleus-nucleus collision zone. 

For K$^+$ the $v_1 (y_{(0)})$ study indicates a minor value
$U_{\text{K}^{+}\text{N}} \approx 10$~MeV. 
Conclusions from the $v_1 (p_{\text{T}}$) study are less clear: 
\begin{itemize}
\item[(a)] 
   within the central collisions, similarly small in-medium effects are favoured; 
\item[(b)]
  for the semi-peripheral collisions, 
  whereas the HSD comparison tends to point to in-medium effect of similar size,
  within the IQMD calculation none of scenarios seem to follow 
  the experimental trend. 
\end{itemize}

In case of K$^-$, the analysis of $v_1 (y_{(0)})$ dependence 
seems to indicate some moderate in-medium effect, which in terms of
$U_{\text{K}^{-}\text{N}}$ could be approximated by about -25~MeV (HSD) 
or about -45~MeV (IQMD).

\end{document}